# Survey of clustering algorithms for MANET

Ratish Agarwal,
[1]Department of Information Technology,
Rajiv Gandhi Proudyogiki Vishwavidyalaya,
Bhopal
ratish@rgtu.net

Dr. Mahesh Motwani
Department of Computer Science Engineering,
Jabalpur Engineering College ,
Jabalpur
mahesh_9@hotmail.com

*Abstract* - **Many clustering schemes have been proposed for ad hoc networks. A systematic classification of these clustering schemes enables one to better understand and make improvements. In mobile ad hoc networks, the movement of the network nodes may quickly change the topology resulting in the increase of the overhead message in topology maintenance. Protocols try to keep the number of nodes in a cluster around a pre-defined threshold to facilitate the optimal operation of the medium access control protocol. The clusterhead election is invoked on-demand, and is aimed to reduce the computation and communication costs. A large variety of approaches for ad hoc clustering have been developed by researchers which focus on different performance metrics. This paper presents a survey of different clustering schemes.**

*Keywords: Mobile ad hoc networks, Clustering, clusterhead, gateway.*

## I. INTRODUCTION

In an ad hoc network, mobile nodes communicate with each other using multihop wireless links. There is no stationary infrastructure; for instance, there are no base stations. Each node in the network also acts as a router, forwarding data packets for other nodes. A research issue in the design of ad hoc networks is the development of dynamic routing protocols that can efficiently find routes between two communicating nodes. The routing protocol must be able to keep up with the high degree of node mobility that often changes the network topology. In a large network, flat routing schemes produce an excessive amount of information that can saturate the network. In addition, given the nodes heterogeneity, nodes may have highly variable amount of resources, and this produces a hierarchy in their roles inside the network. Nodes with large computational and communication power, and powerful batteries are more suitable for supporting the ad hoc network functions (e.g., routing) than other nodes.

Cluster-based routing is a solution to address nodes heterogeneity, and to limit the amount of routing information that propagates inside the network. The idea behind clustering is to group the network nodes into a number of overlapping clusters. Clustering makes possible a hierarchical routing in which paths are recorded between clusters instead of between nodes. This increases the routes lifetime, thus decreasing the amount of routing control overhead. Inside the cluster one node that coordinates the cluster activities is clusterhead (CH). Inside the cluster, there are ordinary nodes also that have direct access only to this one clusterhead, and gateways. Gateways are nodes that can hear two or more clusterheads.

Ordinary nodes send the packets to their clusterhead that either distributes the packets inside the cluster, or (if the destination is outside the cluster) forwards them to a gateway node to be delivered to the other clusters. By replacing the nodes with clusters, existing routing protocols can be directly applied to the network. Only gateways and clusterheads participate in the propagation of routing control/update messages. In dense networks this significantly reduces the routing overhead, thus solving scalability problems for routing algorithms in large ad hoc networks.

## II. CLUSTERING ALGORITHMS IN MANET

We present below a survey of different clustering algorithms.

### 2.1 Identifier-based clustering

A unique ID is assigned to each node. Nodes know the ID of its neighbors and clusterhead is chosen following some certain rules as given below.

**2.1.1 Lowest ID cluster algorithm (LIC)** [1] is an algorithm in which a node with the minimum *id* is chosen as a clusterhead. Thus, the *ids* of the neighbors of the clusterhead will be higher than that of the clusterhead. A node is called a gateway if it lies within the transmission range of two or more clusterheads. Gateway nodes are generally used for routing between clusters. Each node is assigned a distinct *id*. Periodically, the node broadcasts the list of nodes that it can hear (including itself ).

- A node which only hears nodes with *id* higher than itself is a clusterhead.

- The lowest-*id* node that a node hears is its clusterhead, unless the lowest-*id* specifically gives up its role as a clusterhead (deferring to a yet lower *id* node).

- A node which can hear two or more clusterheads is a gateway.

- Otherwise, a node is an ordinary node.

The Lowest-ID scheme concerns only with the lowest node *ids* which are arbitrarily assigned numbers without considering any other qualifications of a node for election as a clusterhead. Since the node *ids* do not change with time, those with smaller *ids* are more likely to become clusterheads than nodes with larger *ids*. Thus, drawback of lowest ID algorithm is that certain nodes are prone to power drainage due to serving as clusterheads for longer periods of time.

**2.1.2 Max-Min d-cluster formation algorithm** [2] generalizes the cluster definition to a collection of nodes that are up to d-hops away from a clusterhead. Due to the large number of nodes involved, it is desirable to let the nodes operate asynchronously. The clock synchronization overhead is avoided, providing additional processing savings. Furthermore, the number of messages sent from each node is limited to a multiple of *d* the maximum number of hops away from the nearest clusterhead, rather than *n* the number of nodes in the network. This guarantees a good controlled message complexity for the algorithm. Additionally, because *d* is an input value to the heuristic, there is control over the number of clusterheads elected or the density of clusterheads in the network. The amount of resources needed at each node is minimal, consisting of four simple rules and two data structures that maintain node information over *2d* rounds of communication. Nodes are candidates to be clusterheads based on their node *id* rather than their degree of connectivity. As the network topology changes slightly the node's degree of connectivity is much more likely to change than the node's *id* relative to its neighboring nodes. If a node *A* is the largest in the d-neighborhood of another node *B* then node *A, A* will be elected a clusterhead, even though node *A* may not be the largest in its d-neighborhood. This provides a smooth exchange of clusterheads rather than an erratic exchange. This method minimizes the amount of data that must be passed from an outgoing clusterhead to a new clusterhead when there is an exchange.

**2.2 Connectivity-based clustering**

**2.2.1 Highest connectivity clustering algorithm (HCC)** [1] The degree of a node is computed based on its distance from others. Each node broadcasts its id to the nodes that are within its transmission range. The node with maximum number of neighbors (i.e., maximum degree) is chosen as a clusterhead. The neighbors of a clusterhead become members of that cluster and can no longer participate in the election process. Since no clusterheads are directly linked, only one clusterhead is allowed per cluster. Any two nodes in a cluster are at most two-hops away since the clusterhead is directly linked to each of its neighbors in the cluster. Basically, each node either becomes a clusterhead or remains an ordinary node.

This system has a low rate of clusterhead change but the throughput is low. Typically, each cluster is assigned some resources which is shared among the members of that cluster. As the number of nodes in a cluster is increased, the throughput drops. The reaffiliation count of nodes is high due to node movements and as a result, the highest-degree node (the current clusterhead) may not be re-elected to be a clusterhead even if it looses one neighbor. All these drawbacks occur because this approach does not have any restriction on the upper bound on the number of nodes in a cluster.

**2.2.2 K-hop connectivity ID clustering algorithm (K-CONID)**[3] combines two clustering algorithms: the Lowest-ID and the Highest-degree heuristics. In order to select clusterheads connectivity is considered as a first criterion and lower ID as a secondary criterion. Using only node connectivity as a criterion causes numerous ties between nodes On the other hand, using only a lower ID criterion generates more clusters than necessary. The purpose is to minimize the number of clusters formed in the network and in this way obtain dominating sets of smaller sizes. Clusters in the K-CONID approach are formed by a clusterhead and all nodes that are at distance at most k-hops from the clusterhead.

At the beginning of the algorithm, a node starts a flooding process in which a clustering request is send to all other nodes. In the Highest-degree heuristic, node degree only measures connectivity for 1-hop clusters. K-CONID generalizes connectivity for a k-hop neighborhood. Thus, when $k = 1$ connectivity is the same as node degree.

Each node in the network is assigned a pair did = (d, ID). d is a node's connectivity and ID is the node's identifier. A node is selected as a clusterhead if it has the highest connectivity. In case of equal connectivity, a node has clusterhead priority if it has lowest ID. The basic idea is that every node broadcasts its clustering decision once all its k-hop neighbors with larger clusterhead priority have done so.

**2.2.3 Adaptive cluster load balance method** [4]. In HCC clustering scheme, one cluster head can be exhausted when it serves too many mobile hosts. It is not desirable and the CH becomes a bottleneck. So a new approach [4] is given. In hello message format, there is an "Option" item. If a sender node is a cluster head, it will set the number of its dominated member nodes as "Option" value. When a sender node is not a cluster head or it is undecided (CH or non-CH), "Option" item will be reset to 0. When a CH's Hello message shows its dominated nodes' number exceeds a threshold (the maximum number one CH can manage), no new node will participate in this cluster. As a result, this can eliminate the CH bottleneck phenomenon and optimize the cluster structure. This algorithm can get load balance between various clusters. Thus, resource consumption and information transmission is distributed to all clusters instead of few clusters.

**2.2.4 Adaptive multihop clustering** [5] is a multihop clustering scheme with load-balancing capabilities. Each mobile node periodically broadcasts information about its ID, Clusterhead ID, and its status (clusterhead/member/gateway) to others within the same cluster. With the help of this broadcast, each mobile node obtains the topology information of its cluster. Each gateway also periodically exchanges information with neighboring gateways in different clusters and reports to its clusterhead. Thus, a clusterhead can know the number of mobile nodes of each neighboring cluster. Adaptive multihop clustering sets upper and lower bounds (U and L) on the number of clustermembers within a cluster that a clusterhead can handle.

When the number of clustermembers in a cluster is less than the lower bound, the cluster needs to merge with one of the neighboring clusters. In order to merge two clusters into one cluster, a clusterhead always has to get the cluster size of all neighboring clusters. It prevents that the number of clustermembers in the merged cluster is over the upper bound. On the contrary, if the number of clustermembers in a cluster is greater than the upper bound, the cluster is divided into two clusters. However, Adaptive multihop clustering does not address how to select a proper node to serve as the clusterhead for the newly detached cluster. The upper and lower bounds are decided by network size, mobility etc.

### 2.3 Mobility-aware clustering

**2.3.1 Mobility-based d-hop clustering algorithm** [6] partitions an ad hoc network into d-hop clusters based on mobility metric. The objective of forming d-hop clusters is to make the cluster diameter more flexible. This algorithm is based on mobility metric and the diameter of a cluster is adaptable with respect to node mobility. This clustering algorithm assumes that each node can measure its received signal strength. In this manner, a node can estimate its distance from its neighbors. Strong received signal strength implies closeness between two nodes. This algorithm requires the calculation of five terms: the estimated distance between nodes, the relative mobility between nodes, the variation of estimated distance over time, the local stability, and the estimated mean distance. Relative mobility corresponds to the difference of the estimated distance of one node with respect to another, at two successive time moments. This parameter indicates if two nodes move away from each other or if they become closer.

The variation of estimated distances between two nodes is computed instead of calculating physical distance between two nodes. This is because physical distance between two nodes is not a precise measure of closeness. For instance, if a node runs out of energy it will transmit packets at low power acting as a distanced node from its physically close neighbor. The variation of estimated distance and the relative mobility between nodes are used to calculate the local stability. Local stability is computed in order to select some nodes as clusterheads. A node may become a clusterhead if it is found to be the most stable node among its neighborhood. Thus, the clusterhead will be the node with the lowest value of local stability among its neighbors.

**2.3.2 Mobility Based Metric for Clustering** [7] proposes a local mobility metric for the cluster formation process such that mobile nodes with low speed relative to their neighbors have the chance to become clusterheads. By calculating the variance of a mobile node's speed relative to each of its neighbors, the aggregate local speed of a mobile node is estimated. A low variance value indicates that this mobile node is relatively less mobile to its neighbors. Consequently, mobile nodes with low variance values in their neighborhoods are chosen as clusterhead. For cluster maintenance, timer is used to reduce the clusterhead change rate by avoiding re-clustering for incidental contacts of two passing clusterheads. However, the mobility behavior of mobile nodes is not always considered in cluster maintenance, so a clusterhead is not guaranteed to bear a low mobility characteristic relative to its members during maintenance phase. As time advances, the mobility criterion is somewhat ignored. This scheme is effective for MANETs with group mobility behavior, in which a group of mobile nodes moves with similar speed and direction, as in highway traffic. Thus, a selected clusterhead can normally promise the low mobility with respect to its member nodes. However, if mobile nodes move randomly the performance may reduce.

**2.3.3 Mobility-based Frame Work for Adaptive Clustering** [8] partition a number of mobile nodes into multi-hop clusters based on (a, t) criteria. The (a, t) criteria indicate that every mobile node in a cluster has a path to every other node that will be available over some time period 't' with a probability 'a' regardless of the hop distance between them. Cluster framework is based on an adaptive architecture designed to dynamically organize mobile nodes into clusters in which the probability of path availability can be bounded, and the impact of routing overhead can be effectively managed. The cluster organization supports an adaptive hybrid routing strategy that is more responsive and effective when node mobility is low and more efficient when node mobility is high. The purpose of this strategy is to support a more scalable routing infrastructure that is able to adapt to high rates of topological change. This is achieved using prediction of the future state of the network links in order to provide a quantitative bound on the availability of paths to cluster destinations. A metric which captures the dynamics of node mobility, makes the scheme adaptive with respect to node mobility.

### 2.4 Low cost of maintenance clustering

**2.4.1 Least cluster change algorithm (LCC)** [9] has a significant improvement over LIC and HCC algorithms as for as the cost of cluster maintenance is consider. Most of protocols executes the clustering procedure periodically, and re-cluster the nodes from time to time in order to satisfy some specific characteristic of clusterheads. In HCC, the clustering scheme is performed periodically to check the "local highest node degree" aspect of a clusterhead. When a clusterhead finds a member node with a higher degree, it is forced to hand over its clusterhead role. This mechanism, involves frequent re-clustering. In LCC the clustering algorithm is divided into two steps: cluster formation and cluster maintenance. The cluster formation simply follows LIC, i.e. initially mobile nodes with the lowest ID in their neighborhoods are chosen as clusterheads. Re-clustering is event-driven and invoked in only two cases:

- When two clusterheads move into the reach range of each other, one gives up the clusterhead role.

- When a mobile node cannot access any clusterhead, it rebuilds the cluster structure for the network according to LIC.

Hence, LCC significantly improves cluster stability by relinquishing the requirement that a clusterhead should HAVE some special features in its local area. But the second case of re-clustering in LCC indicates that a single node's movement may still invoke the complete cluster structure recomputation and thus results in large communication overhead.

**2.4.2 Adaptive clustering for mobile wireless network** [10]. ensures small communication overhead for building clusters because each mobile node broadcasts only one message for the cluster construction. In this adaptive clustering scheme, every mobile node i keep its own ID and the ID of its direct neighbors in a set Gi. Each mobile node with the lowest ID in their local area declares to be a clusterhead and set its own ID as its cluster ID (CID). The CID information includes a mobile node's ID and CID. When a mobile node i receives CID information from a neighbor j, it deletes j from its set Gi. If the CID information from j is a clusterhead claim, the mobile node checks its own CID aspect. If its CID is unspecified (it is not involved in any cluster yet) or larger than the ID (CID) of j, it sets j as its clusterhead. The process continues till all mobile nodes access some cluster. After cluster formation is completed, clusterheads are no longer used in any further cluster maintenance phase. In the maintenance phase, when a mobile node i finds out that the distance between itself and some node j in the same cluster becomes greater than 2-hop, it invokes a cluster maintenance mechanism. If node i is a direct neighbor of the node with the highest intra-cluster connectivity in its cluster, it remains in the cluster and removes node j; otherwise, it joins a neighboring cluster. As soon as there is no proper cluster to join, it forms a new cluster to cover itself. Since this mechanism likely forms new clusters but without any cluster elimination or merge mechanisms, the cluster size decreases and the number of clusters increases as time advances. Eventually, almost every mobile node forms a single-node cluster, and the cluster structure disappears.

**2.4.3 3-hop between adjacent clusterheads (3-hBAC)** [11]. This algorithm introduce a new node status, "clusterguest", which means this node is not within the transmission range of any clusterheads, but within the transmission range of some clustermembers. The cluster formation always begins from the neighborhood of the mobile node with the lowest ID (assuming it is mobile node mo) in a MANET. The mobile node with the highest node degree in mo's closed neighbor set is chosen to be the first clusterhead. All the direct neighbors of this clusterhead change status to "clustermember." After the completion of the first cluster, the cluster formation procedure can be performed in parallel in the network. A clustermember or a direct neighbor of any clustermember with status "unspecified" (indicating it is not included in any cluster yet) are denied serving as a clusterhead. A mobile node, which is not denied clusterhead capability, declares as a new clusterhead when it is with the highest node degree in its neighborhood. When a mobile node finds out that it cannot serve as a clusterhead or join a cluster as a clustermember, but some neighbor is a clustermember of some cluster, it joins the corresponding cluster as a clusterguest.

For cluster maintenance, this algorithm keeps the adjacent clusterheads at least two-hops away. So when two clusterheads move into the reach range of each other, one is required to give up its clusterhead role. With the help of clusterguest, 3hBAC does not raise ripple effect when re-clustering, which means the clusterhead re-election will have no affect on the status of mobile nodes outside these two clusters. For another case, when a mobile node moves out of the ranges of all clusters, it can join a cluster as a clusterguest if it can reach some clustermember(s) of that cluster. Hence, there is no need to form new clusters in order to cover such a single node as in LCC and the cluster topology does not change. This can reduce the number of clusters and eliminate small unnecessary clusters.

**2.4.4 Passive clustering** [12]. Most of the clustering algorithms require all the mobile nodes to announce cluster-dependent information repeatedly to build and maintain the cluster structure, and thus clustering is one of the main sources of control overhead. A clustering protocol that does not use dedicated control packets or signals for clustering specific decision is called Passive Clustering. In this scheme, a mobile node can be in one of the following four states: initial, clusterhead, gateway, and ordinary node. All the mobile nodes are with 'initial' state at the beginning. Only mobile nodes with "initial" state have the potential to be clusterheads. When a potential clusterhead with "initial" state has something to send, such as a flood search, it declares itself as a clusterhead by piggybacking its state in the packet. Neighbors can gain knowledge of the clusterhead claim by monitoring the "cluster state" in the packet, and then record the Cluster head ID and the packet receiving time. A mobile node that receives a claim from just one clusterhead becomes an ordinary node, and a mobile node that hears more claims becomes a gateway. Since passive clustering does not send any explicit clustering-related message to maintain the cluster structure, each node is responsible for updating its own cluster status by keeping a timer. When an ordinary node does not receive any packet from its clusterhead for a given period, its status reverts to "initial".

**2.5 Power-aware clustering**

**2.5.1 Load balancing clustering (LBC)** [13] provide a nearby balance of load on the elected clusterheads. Once a node is elected a clusterhead it is desirable for it to stay as a clusterhead up to some maximum specified amount of time, or budget. The budget is a user defined restriction placed on the algorithm and can be modified to meet the unique characteristics of the system, i.e., the battery life of individual nodes. In this algorithm each mobile node has a variable, virtual ID (VID), and the value of VID is set as its ID number at first. Initially, mobile nodes with the highest IDs in their local area win the clusterhead role. LBC limits the maximum time units that a node can serve as a clusterhead continuously, so when a clusterhead exhausts its duration budget, it resets its VID to 0 and becomes a non-clusterhead node. When two clusterheads move into the reach range, the one with higher VID wins the clusterhead role. when a clusterhead resigns, a non-clusterhead with the largest VID value in the neighborhood can resume the clusterhead function. The newly chosen mobile node is the one whose previous total clusterhead serving time is the shortest in its neighborhood, and this should guarantee good energy level for being a new clusterhead. However, the drawback is that the clusterhead serving time alone may not be a good indicator of energy consumption of a mobile node.

**2.5.2 Power-aware connected dominant set** [14] is an energy-efficient clustering scheme which decreases the size of a dominating set (DS) without impairing its function. The unnecessary mobile nodes are excluded from the dominating set saving their energy consumed for serving as clusterheads. Mobile nodes inside a DS consume more battery energy than those outside a DS because mobile nodes inside the DS bear extra tasks, including routing information update and data packet relay. Hence, it is necessary to minimize the energy consumption of a DS. In this scheme Energy level (el) instead of ID or node degree is used to determine whether a node should serve as clusterhead. A mobile node can be deleted from the DS when its close neighbor set is covered by one or two dominating neighbors, and at the same time it has less residual energy than the dominating neighbors. This scheme cannot balance the great difference of energy consumption between dominating nodes (clusterheads) and non-dominating nodes (ordinary nodes) because its objective is to minimize the DS rather than to balance the energy consumption among all mobile nodes. Hence, mobile nodes in the DS still likely deplete their energy at a much faster rate.

**2.5.3 Clustering for energy conservation** [15] assumes two node types: master and slave. A slave node must be connected to only one master node, and a direct connection between slave nodes is not allowed. Each master node can establish a cluster based on connections to slave nodes. The area of a cluster is determined by the farthest distance between the master node and a slave node in the cluster. Master nodes are s selected in advance, and can only serve a limited number of slave nodes. The purpose of of this scheme is to minimize the transmission energy consumption summed by all master-slave pairs and to serve as many slaves as possible in order to operate the network with longer lifetime and better performance. Two schemes, single-phase clustering and double-phase clustering, are proposed in [15]. In single-phase clustering, initially every master node will page slave nodes with the allowed maximum energy. For each slave that receives one or multiple paging signals, it always sends an acknowledgment message back to the master from which it receives the strongest paging signal. Since a master node can serve only a limited number of slaves, it first allocates channels for slaves that only receive a single paging signal from itself. If any free channels remain, other slave nodes, which receive more than one paging signal, are allocated channels in the order of the power level of the paging signal received from the master node. For those slave nodes, which do not receive a channel from a master in the channel allocation phase, are dropped in the further communication phase. This mechanism can reduce the call drop rate by giving priority to those slave nodes that only receive single paging signals in channel allocation. Slave nodes, which receive multiple paging signals, always try to communicate with the nearest master. Each connected master-slave pair communicates with the minimum transmission power in order to save energy. To further lower the call drop rate, double-phase clustering re-pages for slaves, which do not receive a channel in the first round, in its range. The channel allocation procedure also follows the received signal strength. The drawback of this scheme are paging process before each round of communication consumes a large amount of energy. Master node election is not adaptive, and the method of selecting the master node is not specified.

**2.6 Combined-weight based clustering**

**2.6.1 Weighted clustering algorithm (WCA)** [16] selects a clusterhead according to the number of nodes it can handle, mobility, transmission power and battery power. To avoid communications overhead, this algorithm is not periodic and the clusterhead election procedure is only invoked based on node mobility and when the current dominant set is incapable to cover all the nodes. To ensure that clusterheads will not be over-loaded a pre-defined threshold is used which indicates the number of nodes each clusterhead can ideally support. WCA selects the clusterheads according to the weight value of each node. The weight associated to a node v is defined as:

$$W_v = w_1 \Delta_v + w_2 D_v + w_3 M_v + w_4 P_v \quad \text{---------(1)}$$

The node with the minimum weight is selected as a clusterhead. The weighting factors are chosen so that $w_1 + w_2 + w_3 + w_4 = 1$. $M_v$ is the measure of mobility. It is taken by computing the running average speed of every node during a specified time T. $\Delta_v$ is the degree difference. $\Delta_v$ is obtained by first calculating the number of neighbors of each node. The result of this calculation is defined as the degree of a node v, $d_v$. To ensure load balancing the degree difference $\Delta_v$ is calculated as $|d_v - \delta|$ for every node v, where $\delta$ is a pre-defined threshold. The parameter $D_v$ is defined as the sum of distances from a given node to all its neighbors. This factor is related to energy consumption since more power is needed for larger distance communications. The parameter $P_v$ is the cumulative time of a node being a clusterhead. $P_v$ is a measure of how much battery power has been consumed. A clusterhead consumes more battery than an ordinary node because it has extra responsibilities.

The clusterhead election algorithm finishes once all the nodes become either a clusterhead or a member of a clusterhead. The distance between members of a clusterhead, must be less or equal to the transmission range between them. No two clusterheads can be immediate neighbors

**2.6.2 Entropy-based Weighted clustering algorithm** [17] In WCA high mobility of nodes leads to high frequency of reaffiliation which increase the network overhead. Higher reaffiliation frequency leads to more recalculations of the cluster assignment resulting in increase in communication overhead. Entropy based clustering overcomes the drawback of WCA and forms a more stable network. It uses an entropy-based model for evaluating the route stability in ad hoc networks and electing clusterhead. Entropy presents uncertainty and is a measure of the disorder in a system. So it is a better indicator of the stability and mobility of the ad hoc network.

**2.6.3 Vote-based clustering algorithm** [18] is based on two factors, neighbors' number and remaining battery time of every mobile host (MH) Each MH has a unique identifier (ID) number, which is a positive integer. The basic information inside the network is Hello message, which is transmitted in the common channel. Making use of node location information and power information, this algorithm introduce the concept of "vote". The Hello message format is given below. MH_ID item is MH's own ID and CH_ID item is MH's clusterhead ID, Vote item means MH's vote value, i.e. weighted sum of number of valid neighbors and remaining battery time. Option item is used to realize cluster load balance.

| MH_ID | CH_ID | Vote | Option |
|---|---|---|---|

Hello message format

$$\text{Vote} = w_1 \times (n/N) + w_2 \times (m/M) \quad \text{---------------------------(2)}$$

w1; w2: Weighted coefficient of location factors and battery time, respectively,

n: Number of neighbors,

N: Network size or the Maximum of members in a cluster,

m: Remaining battery time,

M: The maximum of battery time remaining battery time.

Each MH sends a Hello message randomly during a Hello cycle. If a MH is a new user to the network, it reset "CH_ID" item. That means the MH does not belong to any cluster and does not know whether it has neighbor hosts. Each MH counts how many Hello messages it can receive during a Hello period, and considers the number of received Hello messages as its own n. Each MH sends another Hello message, in which "vote" item is set to its own vote value and got from Equation. Recording Hello message during second Hello cycles, each MH knows the sender with highest vote and not belongs to any existing cluster is its cluster head. It set its next sending Hello message item "CH_ID" to the cluster head's ID value. When two or more mobile nodes receive the same number of hello packets, the one who owns the lower ID will get priority. Following this approach, every MH knows its cluster head ID after two Hello message periods.

**2.6.4 Weight-based adaptive clustering algorithm (WBACA)** [19] Drawbacks of WCA algorithm is that all the nodes in the network have to know the weights of all the other nodes before starting the clustering process. This process can take a lot of time. Also, two clusterheads can be one-hop neighbors, which results in the clusters not necessarily being spread out in the network The clustering approach presented in WBACA is based on the availability of position information via a Global Positioning System (GPS). The WBACA considers following parameters of a node for clusterhead selection: transmission power, transmission rate, mobility, battery power and degree. Each node is assigned a weight that indicates its suitability for the clusterhead role. The node with the smallest weight is chosen as the clusterhead. The weight of a node N is defined as:

$$W_N = w_1*M + w_2*B + w_3*T_x + w_4*D + w_5/TR \text{ ------(3)}$$

where wl, w2, w3, w4. and w5 are the weighing factors for the corresponding system parameters listed below: -

M: Mobility of the node

B: Battery power

Tx: Transmission power

D: Degree difference, and

TR: Transmission rate

This algorithm further allows no two clusterheads to be one-hop neighbors of each other. Overlapping clusters are connected through Gateways (nodes connecting two clusterheads). All the ordinary nodes are one-hop from their clusterheads

**2.6.5 Connectivity, energy & mobility driven Weighted clustering algorithm (CEMCA)** [20] The election of the cluster head is based on the combination of several significant metrics such as: the lowest node mobility, the highest node degree, the highest battery energy and the best transmission range. This algorithm is completely distributed and all nodes have the same chance to act as a cluster head. CEMCA is composed of two main stages. The first stage consists in the election of the cluster head and the second stage consists in the grouping of members in a cluster. Normalized value of mobility, degree and energy level is calculated and is used to find the quality (normalized to 1) for each node. The node broadcasts its quality to their neighbors in order to compare the better among them. After this, a node that has the best quality is chosen as a clusterhead. In the second stage the construction of the cluster members set is done. Each clusterhead defines its neighbors at two hops maximum. These nodes form the members of the cluster. Next, each cluster head stores all information about its members, and all nodes record the clusterhead identifier. This exchange of information allows the routing protocol to function in the cluster and between the clusters.

### III. CONCLUSION

We have reviewed several clustering algorithms which help organize mobile ad hoc networks in a hierarchical manner and presented their main characteristics. With this survey we see that a cluster-based MANET has many important issues to examine, such as the cluster structure stability, the control overhead of cluster construction and maintenance, the energy consumption of mobile nodes with different cluster-related status, the traffic load distribution in clusters, and the fairness of serving as clusterheads for a mobile node.

### IV. REFERENCES

[1] M. Gerla and J. T. Tsai, "Multiuser, Mobile, Multimedia Radio Network," *Wireless Networks*, vol. 1, pp. 255–65, Oct. 1995

[2] A.D. Amis, R. Prakash, T.H.P Vuong, D.T. Huynh. "Max-Min D-Cluster Formation in Wireless Ad Hoc Networks". In proceedings of *IEEE Conference on Computer Communications (INFOCOM)* Vol. 1. pp. 32-41, 2000

[3] G. Chen, F. Nocetti, J. Gonzalez, and I. Stojmenovic, "Connectivity based k-hop clustering in wireless networks". Proceedings of the *35th Annual Hawaii International Conference on System Sciences*. Vol. 7, pp. 188.3, 2002

[4] F. Li, S. Zhang, X. Wang, X. Xue, H. Shen, "Vote- Based Clustering Algorithm in Mobile Ad Hoc Networks", In proceedings of *International Conference on Networking Technologies,* 2004

[5] T. Ohta, S. Inoue, and Y. Kakuda, "An Adaptive Multihop Clustering Scheme for Highly Mobile Ad Hoc Networks," in proceedings of *6th ISADS'03*, Apr. 2003

[6] I. Er and W. Seah. "Mobility-based d-hop clustering algorithm for mobile ad hoc networks". *IEEE Wireless Communications and Networking Conference* Vol. 4. pp. 2359-2364, 2004

[7] P. Basu, N. Khan, and T. D. C. Little, "A Mobility Based Metric for Clustering in Mobile Ad Hoc Networks," in proceedings of *IEEE ICDCSW' 01*, pp. 413–18, Apr. 2001

[8] A. B. MaDonald and T. F. Znati, "A Mobility-based Frame Work for Adaptive Clustering in Wireless Ad Hoc Networks," *IEEE JSAC*, vol. 17, pp. 1466–87, Aug. 1999

[9] C.-C. Chiang *et al.*, "Routing in Clustered Multihop, Mobile Wireless Networks with Fading Channel," in proceedings of *IEEE SICON'97*, 1997

[10] C. R. Lin and M. Gerla, "Adaptive Clustering for Mobile Wireless Networks," *IEEE JSAC*, vol. 15, pp. 1265–75, Sept. 1997

[11] J. Y. Yu and P. H. J. Chong, "3hBAC (3-hop between Adjacent Clusterheads): a Novel Non-overlapping Clustering Algorithm for Mobile Ad Hoc Networks," in proceedings of *IEEE Pacrim'03*, vol. 1, pp. 318–21, Aug. 2003

[12] T. J. Kwon *et al.*, "Efficient Flooding with Passive Clustering an Overhead-Free Selective Forward Mechanism for Ad Hoc/Sensor Networks," in proceedings of *IEEE*, vol. 91, no. 8, pp. 1210–20, Aug. 2003

[13] A. D. Amis and R. Prakash, "Load-Balancing Clusters in Wireless Ad Hoc Networks," in proceedings of *3rd IEEE ASSET'00*, pp. 25–32 Mar. 2000

[14] J. Wu *et al.*, "On Calculating Power-Aware Connected Dominating Sets for Efficient Routing in Ad Hoc Wireless Networks," *J. Commun. and Networks*, vol. 4, no. 1, pp. 59–70 Mar. 2002

[15] J.-H. Ryu, S. Song, and D.-H. Cho, "New Clustering Schemes for Energy Conservation in Two-Tiered Mobile Ad-Hoc Networks," in proceedings of *IEEE ICC'01*, vo1. 3, pp. 862–66, June 2001

[16] M. Chatterjee, S. K. Das, and D. Turgut, "An On-Demand Weighted Clustering Algorithm (WCA) for Ad hoc Networks," in proceedings of *IEEE Globecom'00*, pp. 1697–701, 2000

[17] Yu-Xuan Wang, Forrest Sheng Bao, "An Entropy-Based Weighted Clustering Algorithm and Its Optimization for Ad Hoc Networks," wimob,pp.56, *Third IEEE International Conference on Wireless and Mobile Computing, Networking and Communications* (WiMob 2007), 2007

[18] F. Li, S. Zhang, X. Wang, X. Xue, H. Shen, "Vote- Based Clustering Algorithm in Mobile Ad Hoc Networks", proceedings of International Conference on *Networking Technologies,* 2004

[19] S.K. Dhurandher and G.V. Singh" Weight-based adaptive clustering in wireless ad hoc networks" *IEEE* 2005

[20] F.D.Tolba, D. Magoni and P. Lorenz " Connectivity, energy & mobility driven Weighted clustering algorithm " in proceedings of *IEEE GLOBECOM* 2007